\documentclass{Interspeech2024}

\usepackage{hwemoji}
\usepackage{hyperref}
\usepackage{graphicx}
\usepackage{subfigure}
\interspeechcameraready

\title{UTDUSS: UTokyo-SaruLab System for \\ Interspeech2024 Speech Processing Using Discrete Speech Unit Challenge}

\name{$^{*}$Wataru}{Nakata}
\name{$^{*}$Kazuki}{Yamauchi}
\name{Dong}{Yang}
\name{Hiroaki}{Hyodo}
\name{Yuki}{Saito}

\address{
  The University of Tokyo, Hongo 7--3--1, Bunkyo-ku, Tokyo, 113--8656 Japan \\
  $^*$ Equal contribution
  }
\email{nakata-wataru855@g.ecc.u-tokyo.ac.jp, yamauchi-kazuki042@g.ecc.u-tokyo.ac.jp}


\begin{document}

\setlength{\abovedisplayskip}{3pt} 
\setlength{\belowdisplayskip}{3pt} 
\setlength\floatsep{5pt} 
\setlength\intextsep{5pt} 
\setlength\textfloatsep{5pt} 
\setlength{\dbltextfloatsep}{5pt} 
\setlength{\dblfloatsep}{3pt}
\maketitle

\begin{abstract}

We present {\it UTDUSS,} the UTokyo-SaruLab system submitted to Interspeech2024 Speech Processing Using Discrete Speech Unit Challenge.
The challenge focuses on using discrete speech unit learned from large speech corpora for some tasks.
We submitted our UTDUSS system to two text-to-speech tracks: Vocoder and Acoustic+Vocoder.
Our system incorporates neural audio codec (NAC) pre-trained on only speech corpora, which makes the learned codec represent rich acoustic features that are necessary for high-fidelity speech reconstruction.
For the acoustic+vocoder track, we trained an acoustic model based on Transformer encoder-decoder that predicted the pre-trained NAC tokens from text input.
We describe our strategies to build these models, such as data selection, downsampling, and hyper-parameter tuning.
Our system ranked in second and first for the Vocoder and Acoustic+Vocoder tracks, respectively.
\end{abstract}

\section{Introduction}
As the field of neural audio codec (NAC) continues to evolve~\cite{Zeghidour2021SoundStreamAE,kumar2023highfidelity}, its applications have expanded beyond audio encoding/decoding to include audio processing~\cite{Chang2023ExploringSR} and generation~\cite{Borsos2023SoundStormEP,Borsos2022AudioLMAL}.
NAC discretizes input a speech/audio signal that can be utilized for various downstream tasks.
The use of discrete speech units allows for speech processing akin to discrete token processing, leveraging the rich resources developed in the natural language processing field~\cite{Borsos2022AudioLMAL,Wang2023NeuralCL}.
However, speech processing using discrete speech units is a relatively new field, and its full capabilities are yet to be explored.

In light of this, the Interspeech2024 Speech Processing using Discrete Speech Unit Challenge was initiated to encourage research in this area.
The challenge consists of three main tracks: automatic speech recognition (ASR), text-to-speech (TTS), and singing voice synthesis (SVS).
Our team, UTokyo-SaruLab, participated in the TTS track, specifically focusing on two tracks: Vocoder and Acoustic+Vocoder.

This paper presents our system developed for this challenge.
Our approach utilizes a descript audio codec (DAC)-based model for discrete speech unit extraction and vocoding.
Subsequently, a Transformer-based model is employed to construct an acoustic model capable of solving the TTS (Acoustic+Vocoder) task.
Our system, referred to as {\it UTDUSS} (The \textbf{U}niversity of \textbf{T}okyo \textbf{D}iscrete \textbf{U}nit \textbf{S}peech \textbf{S}ynthesizer), achieved the second place in the Vocoder track and the first place in the Acoustic+Vocoder track.
The trained model for the vocoder used in the challenge is available online\footnote{\href{https://huggingface.co/sarulab-speech/UTDUSS-Vocoder}{https://huggingface.co/sarulab-speech/UTDUSS-Vocoder}}.

\section{The Interspeech2024 Speech Processing Using Discrete Speech Unit Challenge}
The challenge comprises three main tracks: ASR, TTS, and SVS.
Our team participated in the TTS track, which is subdivided into two tracks: Vocoder and Acoustic+Vocoder.

{\bf Vocoder:} In this track, participants are provided with a training/development/test split for the Expresso~\cite{nguyen23_interspeech} dataset. The task is to construct a vocoder model which converts discrete speech units to the corresponding waveform. 

{\bf Acoustic+Vocoder:} In this track, participants are provided with a training/development/test split for the LJSpeech~\cite{ljspeech17} dataset.
Two types of training sets are prepared: full training set and 1-hour training set.
The task is to construct an acoustic model which predicts the discrete speech units from an input text and a vocoder model which converts the generated discrete speech units to waveform. Therefore, concatenating the vocoder and acoustic model enables end-to-end TTS.

For both tracks, the primary metrics are UTMOS~\cite{saeki22c_interspeech} and bitrate.
The final ranking is determined by the average rank on these two metrics.
Hence, participants aim to achieve higher UTMOS while reducing bitrate.
For the further details of the challenge, please refer to the \href{https://www.wavlab.org/activities/2024/Interspeech2024-Discrete-Speech-Unit-Challenge/}{challenge page}.

\section{Methods}
In this section, we introduce our methods used for the TTS task.
\subsection{Vocoder track}\label{subsect:vocoder}
In this track, we used DAC~\cite{kumar2023highfidelity} with the number of quantizers set to 2 as the model architecture.
The DAC works as both discrete speech representation extraction and vocoder.
According to the challenge rule, we only used training set of Expresso~\cite{nguyen23_interspeech} for training.
In addition, we applied a few techniques to improve the UTMOS metric.
We denoted the DAC model with all of the following techniques as 😀(pronounced ``smiley'')\footnote{Although we submitted four different systems to the Vocoder track using different settings for codebook size and dataset (☺, 😝
, 😯
, and 😜
), we focus on the best-performing system ``😀'' in this paper.}. 

{\bf Hyper-parameter tuning:}
We noticed that when trained with fewer codebooks, the codebook loss and commitment loss tended to explode.
To overcome this overfitting tendency, we performed manual hyper-parameter tuning on the loss weighting parameters of the DAC training based on the loss values of the training set and development set.

{\bf Matching the model sampling rate to UTMOS:}
The original DAC model configuration assumes 24~kHz and 44.1~kHz sampling rates.
However, the primary metric used for this challenge is UTMOS, which only accepts 16~kHz-sampled speech for the MOS prediction.
Considering that we want to maximize UTMOS with a lower bitrate, modeling with a higher sampling rate seemed to be meaningless in this challenge.
Therefore, we trained our model with a sampling rate of 16~kHz.
    
\begin{figure}
    \centering
    \includegraphics[width=\linewidth]{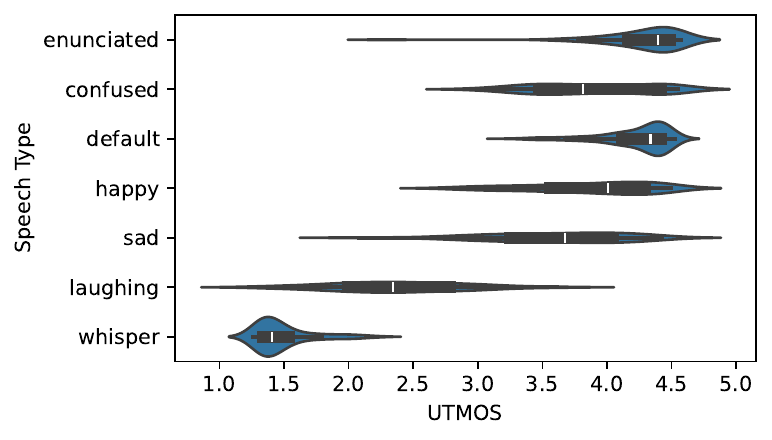}
    \vspace{-8mm}
    \caption{Distribution of UTMOS of the speech in the development subset of EXPRESSO~\cite{nguyen23_interspeech}. We categorize the violin plots by speech types.}
    \label{fig:gt_utmos_dist}
\end{figure}

{\bf Excluding atypical speaking styles from training data:}
Figure~\ref{fig:gt_utmos_dist} shows the violin plots of UTMOS distribution of natural speech samples from the development set used in this challenge.
From this figure, we observe that speech samples with atypical styles covered by EXPRESSO, such as whisper and laughing, tend to have low UTMOS.
One reason is that the UTMOS model was mainly trained on read speech and did not accurately predict naturalness MOS on some expressive speech contained in EXPRESSO.
If these speaking styles are reproduced by the trained DAC model, the UTMOS of synthetic speech can significantly decrease.
Therefore, we tried to make the DAC model generate reading-style speech from discrete speech unit derived from whisper or laughing speech.
To achieve this, we simply excluded these subsets from the training data.

\subsection{Acoustic+Vocoder track}
In this track, we used DAC~\cite{kumar2023highfidelity} with the number of quantizers set to 1 for the discrete speech unit extraction.
As for an acoustic model, we trained a Transformer encoder-decoder model~\cite{vaswani2017attention} that autoregressively predicted DAC tokens from input text, inspired by Transformer TTS~\cite{Li_Liu_Liu_Zhao_Liu_2019}.
As for a vocoder to convert DAC tokens into a waveform, we simply used the decoder in the DAC model used for discrete unit extraction.
We trained the DAC model on the full or 1-hour training set of LJSpeech~\cite{ljspeech17} without any additional datasets.

{\bf Acoustic model based on Transformer encoder-decoder:}
The original Transformer TTS model~\cite{Li_Liu_Liu_Zhao_Liu_2019} takes phoneme sequences as the encoder input and autoregressively generates mel-spectrograms by the decoder through the cross-attention mechanism.
However, autoregressively predicting {\it continuous} mel-spectrograms often suffers from the issue of error accumulation during inference.
In contrast, our acoustic model predicts {\it discrete} speech units (i.e., DAC tokens) in a finite set from a phoneme sequence, which can be expected to mitigate the error accumulation issue better than the original Transformer TTS.

{\bf Sampling strategy:}
Our acoustic model adopts a novel sampling strategy that combines top-$k$~\cite{fan-etal-2018-hierarchical} and top-$p$~\cite{Holtzman2020The} samplings during inference, which can be expected to diversify the output tokens and mitigate the problem of repetition of the same sequence of tokens.
Specifically, to narrow down the candidate tokens, we create a set of the top-$k$ tokens based on the logits (i.e., output distribution of Transformer decoder).
Additionally, we create the smallest set such that the sum of probabilities does not exceed $p$ and low-probability tokens are ignored.
Finally, tokens are sampled from this narrowed-down set according to the probabilities computed from the logits, thereby preserving text diversity while preventing the generation of inappropriate tokens.
This is repeated until the stop token is sampled.
Note that the sampling probability of each token is calculated as a softmax with a temperature parameter applied to the logits.

{\bf Hyper-parameter tuning:} We noticed that the values of the hyper-parameters regarding sampling, i.e., the values of $k$, $p$, and temperature parameter, significantly affected the quality of synthetic speech.
Therefore, we searched for the best values of $k$, $p$, and temperature parameter that maximize UTMOS of synthetic speech.
Specifically, our trained model was input with the transcriptions included in the development set, and the mean of the UTMOS of the generated speech through top-$k$ and top-$p$ sampling was used as the objective for hyper-parameter tuning.
Note that this procedure tunes the hyper-parameters that affect the inference of a pre-trained TTS model, unlike the hyper-parameter tuning in the vocoder track, which tries several training loops with different hyper-parameters.

\section{Experiment}
We conducted experimental evaluations for the techniques applied to our model for each track.

\begin{table*}[t]
    \centering
    \caption{Evaluation results for the TTS (Vocoder) track. {\bf Bold} scores are the best scores among the compared models. Rank shows the rank shown on the leaderboard for the TTS (Vocoder-LowSR) track. Ground truth represents natural speech contained in the test set of Expresso. The baseline was submitted by the challenge organizer (``discrete\_hifigan'').}
    \label{tab:vocoder_ablation}
    \begin{tabular}{l|cccc|c}
    \toprule
    Model & Bitrate ($\downarrow$) & MCD ($\downarrow$) & Log F0 RMSE ($\downarrow$) & UTMOS ($\uparrow$) & Rank \\\midrule
    baseline & \textbf{448.3} & 7.19 & 0.42 & 2.3101 & 12 \\
    DAC (official) & 24046 & {\bf 2.01} & {\bf 0.12} & 3.5596 & - \\\midrule
    😀 & 670 & 4.59 & 0.21 & 3.5815 & {\bf 2} \\
    😀w/o hyper-parameter tuning & 670 & 4.58 & 0.20 & 3.5775 & - \\
    😀w/o data exclusion  & 670 & 4.16 & 0.21 & 3.5679 & - \\
    😀w/o matching sampling rate & 1003 & 4.49 & 0.21 & {\bf 3.6216} & - \\\midrule
    Ground truth &-& -&- & 3.5791 & - \\
    \bottomrule
    \end{tabular}
\end{table*}

\begin{table*}[tb]
\centering
\caption{Evaluation results for the TTS (Acoustic+Vocoder) track. {\bf Bold} scores are the best scores among the compared models. Rank shows the ranks shown on the leaderboard for the TTS (Acoustic+Vocoder) track. Ground truth represents natural speech contained in the test set of LJSpeech. The baseline was submitted by the challenge organizer (``fastspeech2'').}
\vspace{-3mm}
\label{tab:tts_result}
\subtable[Train on the full training set of LJSpeech]{
\label{tb:tts_resulta}
    \begin{tabular}{l|c|ccccc|c}
    \toprule
    Model & Codebook size & Bitrate ($\downarrow$) & MCD ($\downarrow$) & Log F0 RMSE ($\downarrow$) & WER ($\downarrow$) & UTMOS ($\uparrow$) & Rank \\\midrule
    baseline & - & 448.3 & 7.19 & \textbf{0.26} & 8.1 & 3.73 & 9 \\\midrule
    DiscreteTTS-v1.2 & 1024 & 351.1 & 7.15 & 0.29 & 7.5 & 4.2947 & 8 \\
    DiscreteTTS-v2.2 & 512 & 313.8 & {\bf 4.36} & 0.29 & {\bf 7.2} & {\bf 4.3597} & {\bf 1} \\
    DiscreteTTS-v3 & 256 & {\bf 277.6} & 6.96 & 0.29 & 7.7 & 4.3344 & 2 \\\midrule
    Ground truth & - & - & - & - & - & 4.4345 & - \\
    \bottomrule
    \end{tabular}
}
\subtable[Train on the 1-hour training set of LJSpeech]{
\label{tb:tts_resultb}
    \begin{tabular}{l|c|ccccc|c}
    \toprule
    Model & Codebook size & Bitrate ($\downarrow$) & MCD ($\downarrow$) & Log F0 RMSE ($\downarrow$) & WER ($\downarrow$) & UTMOS ($\uparrow$) & Rank \\\midrule
    baseline & - & 448.3 & \textbf{7.55} & \textbf{0.33} & \textbf{7.3} & \textbf{2.4757} & 4 \\\midrule
    DiscreteTTS-v1.2 & 1024 & 331.6 & 11.31 & 0.35 & 110.4 & 2.2613 & 6 \\
    DiscreteTTS-v2.2 & 512 & 313.8 & 11.32 & 0.35 & 111.8 & 2.3607 & {\bf 1} \\
    DiscreteTTS-v3 & 256 & {\bf 267.6} & 11.33 & 0.36 & 112.9 & 2.2896 & 2 \\\midrule
    Ground truth & - & - & - & - & - & 4.4345 & - \\
    \bottomrule
    \end{tabular}
}
\end{table*}
\begin{table}[t]
    \centering
    \caption{Best parameters obtained by hyper-parameter tuning in TTS track.}
    \vspace{-3mm}
    \label{tab:tts_best-params}
    \begin{tabular}{c|ccc}
    \toprule
    codebook size & $k$ & $p$ & temperature \\\midrule
    1024 & 11 & 0.186 & 0.507 \\
    512 & 176 & 0.521 & 0.375 \\
    256 & 181 & 0.779 & 0.351 \\
    \bottomrule
    \end{tabular}
\end{table}

\begin{figure*}[t]
  \centering
  \includegraphics[width=0.8\linewidth]{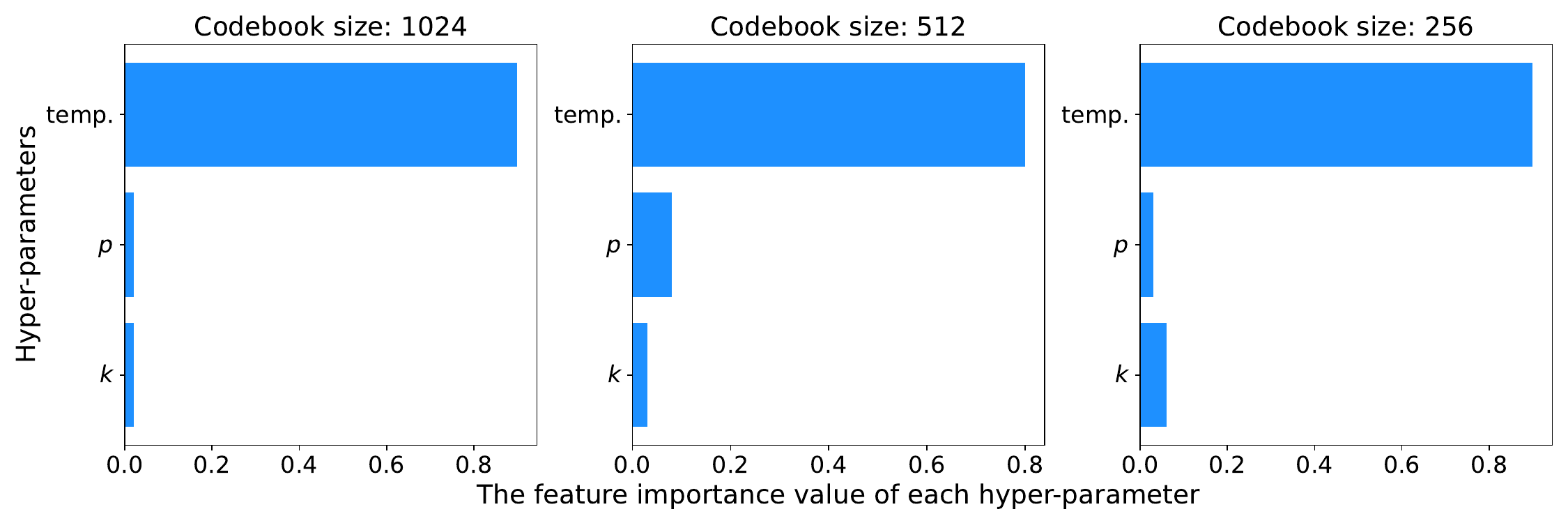}
  \vspace{-4mm}
  \caption{The feature importance value of the values of $k$, $p$, and temperature (temp. in the figure) parameter evaluated by Optuna.}
  \label{fig:importance}
\end{figure*}

\begin{figure*}[t]
  \centering
  \includegraphics[width=\linewidth]{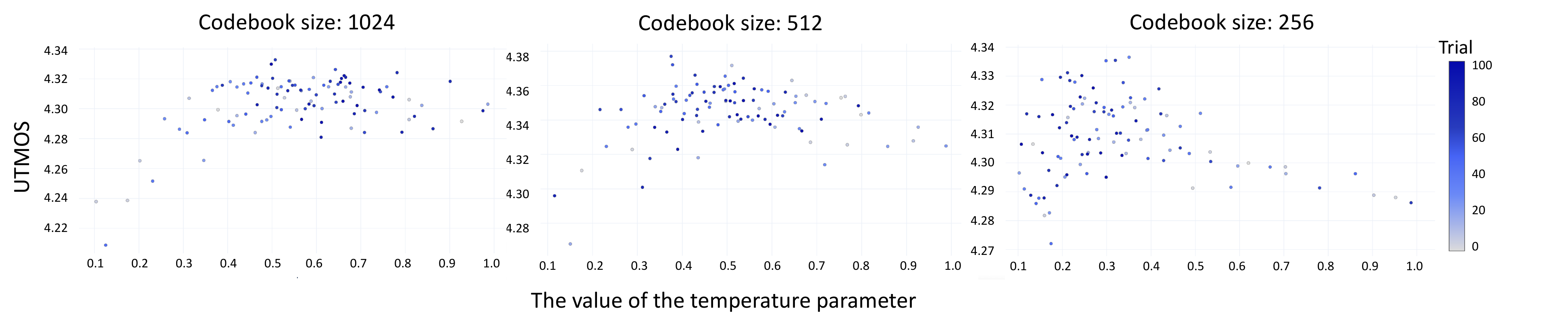}
  \vspace{-8mm}
  \caption{Distribution of UTMOS with respect to the value of the temperature parameter in hyper-parameter tuning.}
  \label{fig:temperature}
\end{figure*}

\subsection{Vocoder track}

We compared the following models.
\begin{itemize}
    \item {\bf DAC (official):} DAC with the official implementation and model parameters distributed on Github. The DAC model was trained on 24~kHz-sampled speech.
    \item {\bf 😀:} DAC with the all proposed techniques described in Section~\ref{subsect:vocoder}.
    \item {\bf 😀w/o hyper-parameter tuning:} This proposed model was trained with the default loss weighting parameters (i.e., 0.25 for the commitment loss of RVQ and 1.0 for the codebook loss).
    \item {\bf 😀w/o data exclusion:} This proposed model was trained on the full training set of the Expresso dataset (i.e., including whisper and laughing subsets).
    \item {\bf 😀w/o matching sampling rate:} This proposed model was trained on 24~kHz-sampled speech.
\end{itemize}

{\bf Dataset:}
We used the Expresso dataset following the training/development/test split provided by the challenge organizers. 
When training ``😀w/o data exclusion,'' we used full training set. Other models were trained without the laughing and whisper speech types.
All audio were resampled from 48~kHz to 16~kHz except for the ``😀w/o matching sampling rate.''

{\bf Experimental setup:}
For model configuration, we set the DAC encoder downsampling rates to $2,3,4,4,5$ and $5,4,4,3,2$ achieving downsampling rate of 480.
When trained on 16~kHz-sampled and 24~kHz-sampled speech, the model achieved discrete speech unit output frequency of 100/3~Hz and 50~Hz, respectively.
For the multi-period discriminator, we set the periods to $2,3,5,7,11,13,17$ and FFT window sizes to $2048,1024,512,256$.
The codebook size (i.e., the number of code vectors in the codebook) was 1024 and the number of codebook in residual vector quantization (RVQ)~\cite{vasuki06} was set to 2.
For loss weighting, we set the weight on commitment loss of RVQ to 2.0 and codebook loss to 8.0, which were tuned on the basis of hyper-parameter search and close to the weight for the mel-spectrogram prediction loss (15.0).
For optimization, we set the batch size to 128 distributed across eight NVIDIA A100 GPUs with automatic mixed precision.
The duration of each training was 1 second for ``😀w/o matching sampling rate'' and 1.5 seconds for other models.
The training was performed for 230k steps in approximately 30 hours.
For other hyper-parameters, we followed the configuration of the \href{https://github.com/descriptinc/descript-audio-codec/blob/main/conf/final/24khz.yml}{official model}.

{\bf Objective evaluation:}
We performed objective evaluations using UTMOS, bitrate, mel-cepstral distortion (MCD), and Log F0 root-mean-square error (RMSE), following the metrics used for the discrete speech unit challenge. 
For calculation of each metric, please refer to the \href{https://www.wavlab.org/activities/2024/Interspeech2024-Discrete-Speech-Unit-Challenge/}{challenge page}.


\subsection{Acoustic+Vocoder track}

We prepared a DAC model with the number of codebook in RVQ set to 1 and codebook size of 1024, 512, and 256.
The other settings were the same as those of the DAC model for the vocoder track.
We denoted the TTS (Acoustic+Vocoder) model with the DAC codebook size of 1024, 512, and 256 as DiscreteTTS-v1.2, DiscreteTTS-v2.2, and DiscreteTTS-v3, respectively and submitted them.

{\bf Dataset:}
We used the LJSpeech dataset for training the TTS model, following the training/development/test split provided by the challenge organizers.
We prepared models trained on the full or 1-hour training set of LJSpeech.
All audio were resampled to 16~kHz.

{\bf Experimental setup:}
We mainly followed the official implementation of \href{https://github.com/soobinseo/Transformer-TTS}{Transformer~TTS} for the network architecture and training settings.
The acoustic model was trained on the full training set with a batch size of 32, learning rate of 0.001, and 200k iterations in approximately 3 hours, while on the 1-hour training set with a batch size of 32, learning rate of 0.0001, and 100k iterations in approximately 2 hours.
We used a single NVIDIA A100 GPU and the Adam optimizer~\cite{kingma14adam}.
Also, we searched for the best hyper-parameters for each model using \href{https://github.com/optuna/optuna}{Optuna}~\cite{optuna}.
The values of $p$ and the temperature parameter were searched within the range of 0.1 to 1.0.
The value of $k$ was searched within the range of 5 to 300 for codebook sizes of 1024 and 512, and within the range of 5 to 200 for codebook size of 256.
Hyper-parameter tuning was performed on only models trained with the full training set.

{\bf Objective evaluation:}
We performed objective evaluations using the UTMOS, bitrate, MCD, LogF0 RMSE, and word error rate (WER) as the evaluation criteria, following the discrete speech unit challenge. 
For calculation of each metric, please refer to the \href{https://www.wavlab.org/activities/2024/Interspeech2024-Discrete-Speech-Unit-Challenge/}{challenge page}.


\section{Results and Discussion}
\subsection{Vocoder track}
Table~\ref{tab:vocoder_ablation} shows the evaluation results for the techniques used in the challenge.
From this result, we observe that our proposed models significantly outperform the baseline in terms of UTMOS, demonstrating the effectiveness of the use of DAC as the discrete token extractor and vocoder.
Especially, the data exclusion and the hyper-parameter tuning strategies are effective on this track to improve the UTMOS metric.
However, from the results of ``😀w/o matching sampling rate,'' the downsampling to 16~kHz worsens the UTMOS metric. 
One reason is that ``😀w/o matching sampling rate'' has the same output frequency as wav2vec~2.0~\cite{baevski2020wav2vec2}, i.e., the backbone self-supervised learning model used for the UTMOS prediction.
However, ``😀w/o matching sampling rate'' is the worst among the proposed models in terms of bitrate. 
For the submitted model, we selected ``😀'' because it has a good balance of the UTMOS and bitrate metrics.
When compared against ``DAC (official),'' all of our models are better in both bitrate and UTMOS metrics.
This is because while official DAC model is trained on various audio data, our models are specialized in the Expresso dataset.
Hence, when evaluated on speech from Expresso, our model performs better.

\subsection{Acoustic+Vocoder track}

Table~\ref{tab:tts_result} shows the evaluation results for our TTS models.
For both the full training and the 1-hour training, our models with codebook size of 512 ranked in first regarding the trade-off between UTMOS and bitrate.
For the full training, our models outperformed the baseline both in terms of UTMOS and bitrate, and the synthetic speech achieved UTMOS close to that of natural speech.
In contrast, for the 1-hour training, UTMOS of the synthetic speech significantly degraded and WER became very high, underperforming the baseline in terms of quality of synthetic speech.
This is mainly due to overfitting to the training data.
While the output speech sounded like the target speaker's voice, the spoken content was inconsistent with the given transcription.

Table~\ref{tab:tts_best-params} shows the results of hyper-parameter tuning.
From these results, we observe that for a codebook size of 1024, smaller values of $k$ and $p$, i.e., reducing the number of candidate DAC tokens during sampling, lead to higher UTMOS on the development set. Additionally, as the codebook size decreases, the best values of $k$ and $p$ tend to increase, while the temperature parameter tends to decrease.
Fig.~\ref{fig:importance} shows the importance of each hyper-parameter evaluated based on completed trials during hyper-parameter tuning by Optuna.
From this figure, we confirm that the temperature parameter is very important compared to the values of $k$ and $p$.
Also, Fig.~\ref{fig:temperature} shows the distribution of UTMOS with respect to the temperature parameter values in hyper-parameter tuning.
This figure illustrates that the UTMOS distributions are indeed affected by the settings of the temperature parameter and codebook size.
Specifically, there is a tendency for the range of temperature parameter values with high UTMOS to become lower as the codebook size decreases.
There results suggest that tuning the sampling strategy is crucial for discrete speech unit-based TTS achieving higher UTMOS while reducing bitrate.

\section{Conclusion}
In this paper, we introduced {\it UTDUSS}, our system for Interspeech2024 speech processing using discrete speech unit challenge.
UTDUSS ranked first on the Acoustic+Vocoder track for both the full and 1-hour training and second in the Vocoder track.
The evaluation results were performed to see the effectiveness of the techniques used in our systems.
Future work includes the better discrete modeling for speech synthesis, including TTS as well as voice conversion.

\section{Acknowledgement}
This work was supported by JST, Moonshot R\&D Grant Number JPMJPS2011.

\bibliographystyle{IEEEtran}
\bibliography{mybib}

\end{document}